\documentclass[final,5p,times,twocolumn,compress]{elsarticle}
\usepackage{amssymb}
\usepackage{amsmath}
\usepackage{graphicx}
\usepackage{float}
\usepackage{dcolumn}
\usepackage{multirow}
\usepackage{changes}
\usepackage[utf8]{inputenc}
\usepackage[T1]{fontenc}
\usepackage[colorlinks = true,linkcolor = blue,urlcolor  = blue,citecolor = blue,anchorcolor = blue]{hyperref}
\usepackage{enumerate}
\usepackage{times}
\usepackage{microtype}
\usepackage{orcidlink}
\usepackage[percent]{overpic}
\usepackage{stfloats}
\usepackage[colorlinks = true,linkcolor = blue,urlcolor  = blue,citecolor = blue,anchorcolor = blue]{hyperref}
\usepackage{caption}
\usepackage{bm}
\usepackage{svrsymbols}

\journal{Physics Letters B}
\begin{document}
\begin{frontmatter}


\title{Implications of the recent neutron decay measurements \\ on the properties of compact objects - a dark star with nucleonic shell ?}

\author[first]{M. Veselsk\'y\orcidlink{0000-0002-7803-0109}}
\ead{Martin.Veselsky@cvut.cz}

\author[first]{V. Petousis\orcidlink{0000-0002-5575-6476}}
\ead{vlasios.petousis@cvut.cz}

\author[second]{Ch.C. Moustakidis\orcidlink{0000-0003-3380-5131}}
\ead{moustaki@auth.gr}

\author[second]{M. Vikiaris\orcidlink{0009-0005-8779-4556}}
\ead{mvikiari@auth.gr}

\affiliation[first]{organization={Institute of Experimental and Applied Physics, Czech Technical University},
            city={Prague},
            postcode={11000}, 
            country={Czechia}}

\affiliation[second]{organization={Department of Theoretical Physics, Aristotle University of Thessaloniki},
            city={Thessaloniki},
            postcode={54124}, 
            country={Greece}}

\begin{abstract}
Recent experimental observation suggests that neutron decay is always accompanied by emission of electron while in 1\% of cases proton is not emitted. We develop a scenario kinematically compatible with experimental observation, where neutron decay results in production of two dark matter particles of about half the mass of neutron and test properties of neutron stars with admixture of such particles. Constraints on mass and coupling to vector dark boson are obtained. The structure of the compact object is modified to a dark star with a shell of nucleonic matter around the nuclear saturation density.
\end{abstract}

\end{frontmatter}

\section{Introduction}

Neutron stars (NSs) serve as natural laboratories for fundamental physics, where extreme densities and strong gravitational fields provide a unique environment to probe nuclear interactions and particle physics beyond terrestrial experiments. One of the unresolved mysteries in particle physics is the "neutron decay puzzle", where discrepancies between the neutron lifetime measured in beam and bottle experiments suggest the potential involvement of new physics, such as exotic decay channels or interactions with dark matter (DM). These inconsistencies in neutron decay rates could have profound implications for NS structure, stability, and evolution. 

Following the hypothesis in a series of papers ~\cite{Fornal-2020, Motta-2018a,Motta-2018b,Husain-2022a,Husain-2023,Husain-2025,Shirke-2023,Das-Burgio-2025,Tan-2019,Cline-2018,Husain-2022b,Darini-2023,Strumia-2022,Ejiri-2019,Ivanov-2019}, the effect of the dark neutron decay on the properties of neutron stars and finite nuclei was studied.
Given that NSs are composed predominantly of degenerate neutrons, even slight modifications to neutron decay properties can influence their composition, equation of state (EOS), and cooling mechanisms. Additionally, if exotic decay modes are realized in NSs, they may alter the balance of beta equilibrium, affecting the proton fraction and the possibility of direct Urca processes, which significantly impact the thermal evolution of NSs.

In this letter, we explore the theoretical consequences of the neutron decay puzzle on NS properties in the light of recent experimental progress. We analyze potential modifications to the EOS, changes in mass-radius relations, and possible effects on NS cooling. Furthermore, we discuss observational signatures that could hint at deviations from the standard neutron decay paradigm, providing a possible astrophysical probe for new physics.

This letter is organized as follows: in section 2, we present the theoretical model, in section 3 we provide our results with a discussion and the section 4 contains our conclusions of this work.

\begin{table}[b]
\caption{Parameter set for RMF EOS \cite{Kanakis-Petousis-2024,Kaons1}.}
\begin{center}
\vspace{0.1cm}
    \begin{tabular}{lrr}
    \hline
    Parameters &  & Units \\
    \hline
        m$_{\omega}$ & 783.00 & MeV \\
        m$_{\sigma}$ & 508.57 & MeV \\
        g$_{\omega N}$ & 9.52 & \\
        g$_{\sigma N}$ & 8.48 & \\
        g$_{\rho N}$  & 4.69 &  \\
        $\kappa$ & 37.18 & MeV \\
        $\lambda$ & -146.86 &  \\
    \hline
    \end{tabular}
\end{center}
\label{tab:tabeos}
\end{table}

\begin{figure*}[t!]
\begin{center}
\includegraphics[width=\columnwidth]{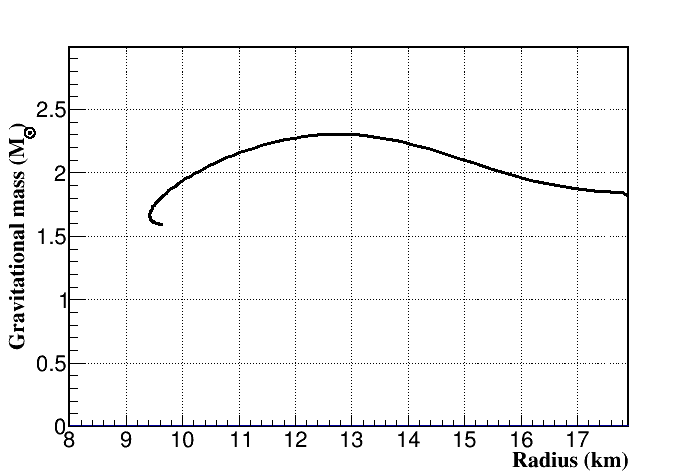}
~
\includegraphics[width=\columnwidth]{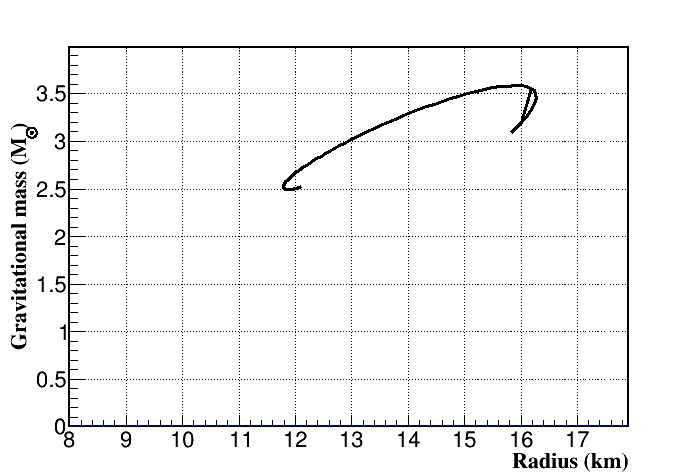}
~
\includegraphics[width=\columnwidth]{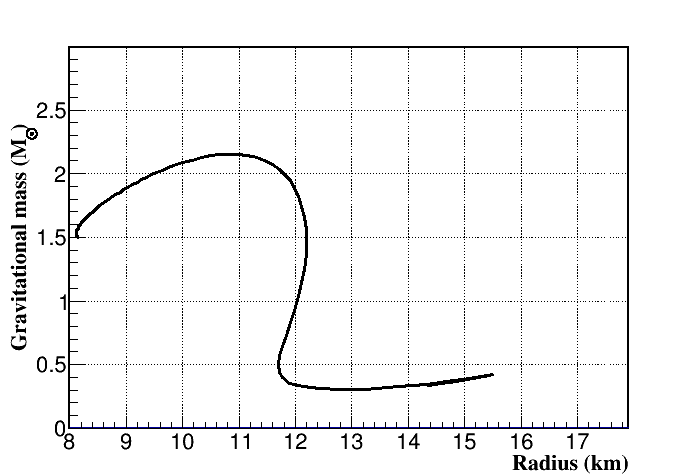}
~
\includegraphics[width=\columnwidth]{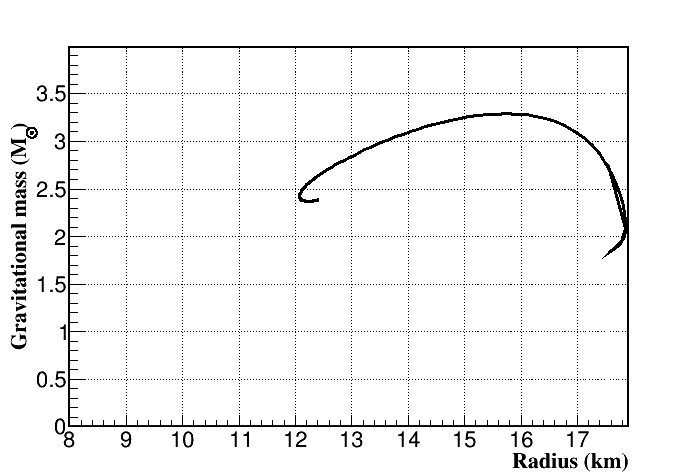}
~
\includegraphics[width=\columnwidth]{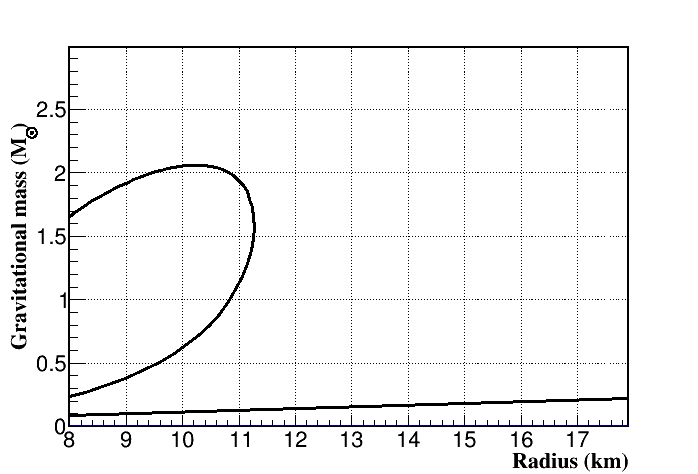}
~
\includegraphics[width=\columnwidth]{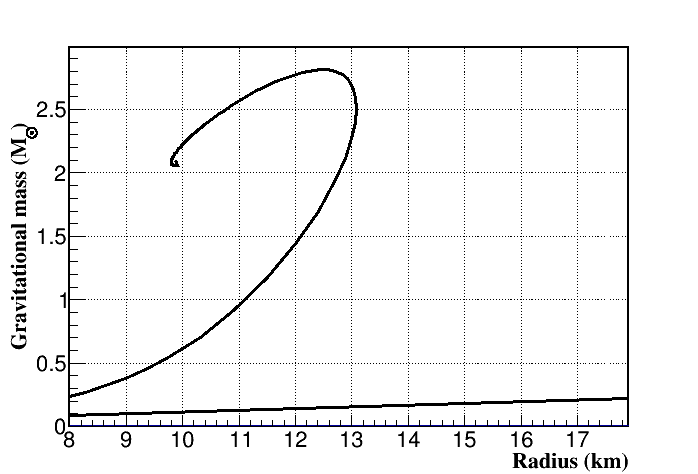}

\caption{Mass-radius plots for several values of m$_D$ and  $g_{vD}$/$m_{vD}$. (upper row) left - m$_D$=340 MeV,  $g_{vD}$/$m_{vD}$=0.015 MeV$^{-1}$, right - m$_D$=340 MeV, $g_{vD}$/$m_{vD}$=0.065 MeV$^{-1}$, (middle row) left - m$_D$=400 MeV, $g_{vD}$/$m_{vD}$=0.015 MeV$^{-1}$, right - m$_D$=400 MeV, $g_{vD}$/$m_{vD}$=0.095 MeV$^{-1}$, (bottom row) left - m$_D$=460 MeV, $g_{vD}$/$m_{vD}$=0.015 MeV$^{-1}$, right - m$_D$=460 MeV, $g_{vD}$/$m_{vD}$=0.095 MeV$^{-1}$.}
\label{fig1}
\end{center}
\end{figure*}

\section{The theoretical model}

The so-called neutron decay puzzle consists of different results of neutron half-life measurements when using different methods. In the so-called bottle experiments \cite{ndbottle1,ndbottle2,ndbottle3,ndbottle4,ndbottle5,ndbottle6,ndbottle7,ndbottle8} cold-neutrons loss is being counted and the resulting half-life is 878.4$\pm$0.5 s. In the beam experiments \cite{ndbeam1,ndbeam2} the protons as remnants of beta-decay are counted and the measured half-life is longer by 10s, amounting to 888.0$\pm$2.0 s. This discrepancy leads to explanations by theorists as a possible decay into electrically neutral DM particles, most prominently as decay into one heavy dark fermion and one light dark boson \cite{Bartosz,Baym,McKeen,Grinstein, Gil}. Recent experimental work at J-PARC \cite{2412.19519} attempted to cure the discrepancy and using the beam method counted electrons. 

In this case the measured half-life is again 877.2$\pm$1.7 s as in the bottle method. This appears to mean that in each neutron decay one electron is produced and if one considers also the results with counting protons as correct then eventuality emerges of neutron decay where proton is not produced while electron is emitted. This would represent a modified neutron puzzle and mean that standard model baryon number is not conserved and at least one further product of decay must be charged what is in conflict with common assumption that DM must be electrically neutral. The question is why such charged particle is not seen in experiment. If e.g. electron-positron pair would be created simultaneously, kinematic properties of both electron and positron would be similar and both particles would have to be detected in the experiment at J-PARC. 

Only explanation is that the kinematic properties of the missing charge are such that it does not fit into experimental angular and momentum acceptance or becomes indistinguishable from background. This would suggest a multi-step process where e.g. neutron first undergoes a process analogous to beta-decay, producing short-lived dark charged fermion, electron and neutrino with similar kinematic properties as electrons and neutrinos from regular neutron beta-decay: \( n^{o} \rightarrow X^{+}+e^{-}+\antineutrino_{e}\) and then the charged dark fermionic remnant (here noted as $X^{+}$) immediately undergoes decay into two (or more) equal mass neutral particles (here noted as $A^{0}$) and light positively charged particle like positron: \( X^{+} \rightarrow 2A^{0}+e^{+}\) which in such 3-body decay will carry enough transverse momentum to break the kinematic correlation to detected electron. The positively charged positron and two (or more) approximately equal mass mid-heavy dark particles will thus be produced. These equal mass particles can be either fermions or bosons.

Obviously baryonic number is not conserved in such neutron decay but that can be expected in this case. Of these particles, energetic positron is in principle observable, but due to its high energy will leave less visible track, especially when using inert helium gas as a working medium and thus can go unobserved. Also such particle with wide angular distribution can be considered as a part of the background caused by cosmic muons. 
When considering implications of such neutron decay for NSs, one can formulate conditions of chemical equilibrium in such a way that: 

\begin{equation}
\mu_{n} = n \times \mu_D + \mu_{e^-}
\end{equation}

where n = 2 or more. The positron produced in the dark neutron decay will ultimately annihilate with one of the electrons in the matter and no positron fraction and thus also no corresponding chemical potential need to be considered. Along with this, also standard condition for beta-equilibrium should be fulfilled: 

\begin{equation}
\mu_n - \mu_p = \mu_{e^-}
\end{equation}

and in this case the matter should be treated as a single fluid. Besides $\beta$-equilibrium, also charge neutrality is required. 

\begin{minipage}[t]{0.48\textwidth}
  \includegraphics[width=\linewidth] {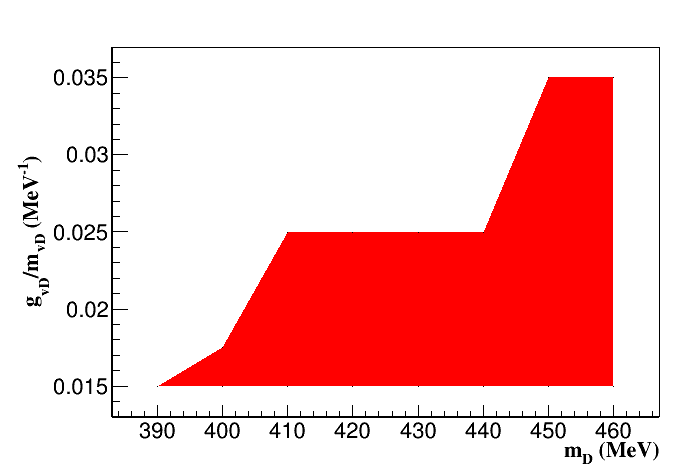}
\captionof{figure}{Combinations of DM particle mass m$_D$ and  $g_{vD}$/$m_{vD}$ fulfilling the NS constraints.}
\label{fig2}
\end{minipage}

\begin{figure*}[t!]
\includegraphics[width=\columnwidth]{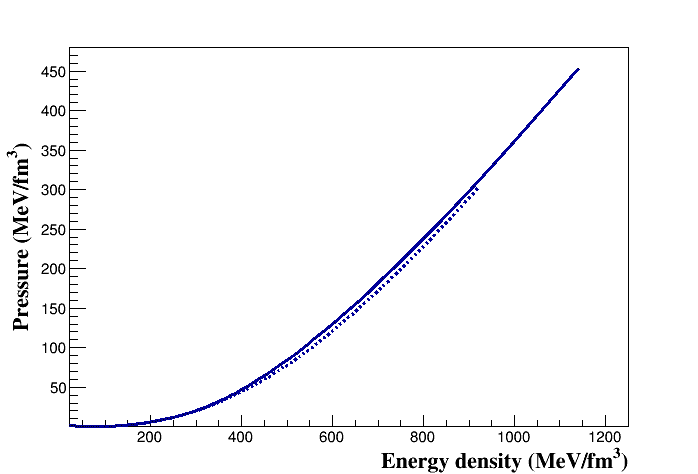}
~
\includegraphics[width=\columnwidth]{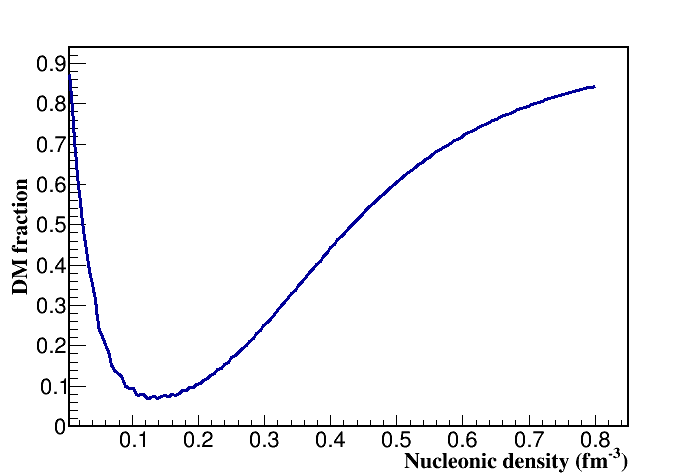}
\caption{(Left) Equation of state with DM particle mass m$_D$=400 MeV and $g_{vD}$/$m_{vD}$=0.015 MeV$^{-1}$(solid line) compared to initial EOS from Table 1 (dashed line), (Right) DM particle fraction for m$_D$=400 MeV, $g_{vD}$/$m_{vD}$=0.015 MeV$^{-1}$.}
\label{fig3}
\end{figure*}

The nucleonic EOS can be chosen requiring to fulfill usual constraints such as allowing to exceed maximum NS mass of 2 $M_{\odot}$ and fit within the range of radius at 1.4 $M_{\odot}$ within constraints given by recent measurements by NICER or extracted from the observation of gravitational wave signal from GW170817 binary NS merger \cite{Abbott}. In particular, we chose the EOS obtained using relativistic mean field (RMF) model with the parameters shown in Table 1 and used in \cite{Kanakis-Petousis-2024}. In simplest variant, the fermionic DM particles can be represented as zero temperature non-interacting Fermi gas and its fraction can be determined using conditions of chemical equilibrium. However, this simple picture is usually insufficient and repulsive interaction mediated by dark vector boson is also considered. In the manner consistent with RMF model, energy density and pressure of DM is expressed as:
\begin{align}
{\cal E}_{DM}&=\frac{(\hbar c)^3g_{vD}^2}{2(m_{vD}c^2)^2}n_D^2+\frac{\gamma}{(2\pi)^3}\int_0^{k_{FD}} 4\pi k^2 \sqrt{(\hbar c k)^2+(m_D c^2)^2}dk
 \label{RMF-E-1}
\end{align}
\begin{align}
{\cal P}_{DM}&=\frac{(\hbar c)^3g_{vD}^2}{2(m_{vD}c^2)^2}n_D^2+\frac{1}{3}\frac{\gamma}{(2\pi)^3}\int_0^{k_{FD}}\frac{4\pi k^2}{\sqrt{(\hbar c k)^2+(m_D c^2)^2}}dk
 \label{RMF-P-1}
\end{align}

Here ${\cal E}_{DM}$ is the DM energy density, ${\cal P}_{DM}$ is the DM pressure, $g_{vD}$ is coupling of the vector boson to DM, $m_{vD}$ is the rest masses of vector DM boson, $m_D$ is the rest mass of the DM particle.
In addition, $n_D$ is the DM density, $k_{FD}$ is the Fermi momentum of DM particle at zero temperature and $\gamma$ is the degeneracy with value $\gamma= 2$. 

\section{Results and discussion}

The strength of the coupling relative to dark vector boson mass $g_{vD}$/$m_{vD}$ is being varied between zero and 0.1 MeV$^{-2}$, thus exceeding typical value 0.02 MeV$^{-2}$ for nucleonic matter. Mass of the DM particle is limited from above by half the nucleon mass. 
The resulting EOS for given DM particle mass and coupling is then used to solve Tolman-Oppenheimer-Volkov (TOV) equation. The resulting mass-radius plots for several cases are shown in Figure \ref{fig1}. Variety of dependencies can be observed, with maximum mass dropping with increasing DM particle mass. The properties of eventual low mass objects for given DM particle mass are influenced by the coupling to vector field. Of the investigated range of DM particle masses 300 – 460 MeV and coupling $g_{vD}$/$m_{vD}$ from 0 to 0.1, only mass range of 390 – 460 MeV and coupling range 0.015 – 0.035 MeV$^{-1}$ (see Figure \ref{fig2}) lead to mass-radius dependencies which fulfill the maximum NS mass of 2 $M_{\odot}$ while remaining compatible with the radius constraints for a range of radius at 1.4 $M_{\odot}$ determined from recent NICER measurements and from the extracted gravitational-wave signal of the binary NS merger GW170817.
Thus a rather strong constraints on the DM particle properties are obtained. Also it is worth noting that simplest variant without such coupling does not fulfill the NS constraints and it is thus excluded. In the Figure \ref{fig3} is shown the EOS and fraction of DM particles as a function of energy density for the successful case of m$_D$=400 MeV and $g_{vD}$/$m_{vD}$=0.015 MeV$^{-1}$. The EOS is practically identical to initial nucleonic EOS, thus suggesting that the dark fermion could be dark analogue of nucleon. Also the value of $g_{vD}$/$m_{vD}$, which is close to the value typical for nucleus, supports such conclusion. On the other hand, such equivalence could prevent possibility to distinguish such star from regular neutron stars by atrophysical observations. 

The plot of DM particle fraction shows that DM particle practically replaces nucleons at low densities and its fraction drops with increasing energy density until reaching minimum around nuclear saturation density, then increases again and at high densities appears to replace nucleonic matter again. The compact object would be then formed mostly by fermionic DM matter with a shell of nucleonic matter around the interface between crust and core. 

\begin{figure*}[t!]
\includegraphics[width=\columnwidth,height=0.65\columnwidth]{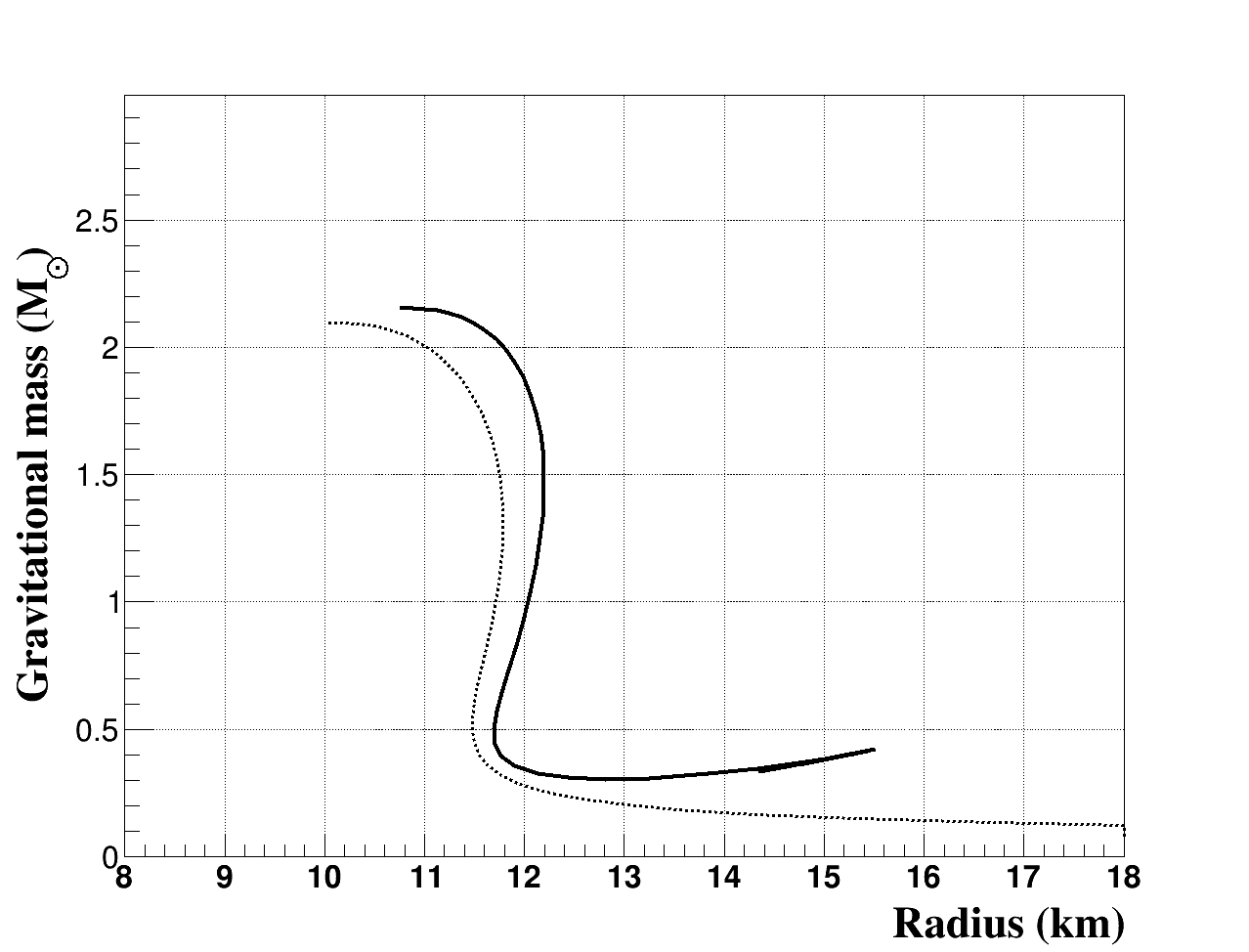}
~
\includegraphics[width=\columnwidth,height=0.65\columnwidth]{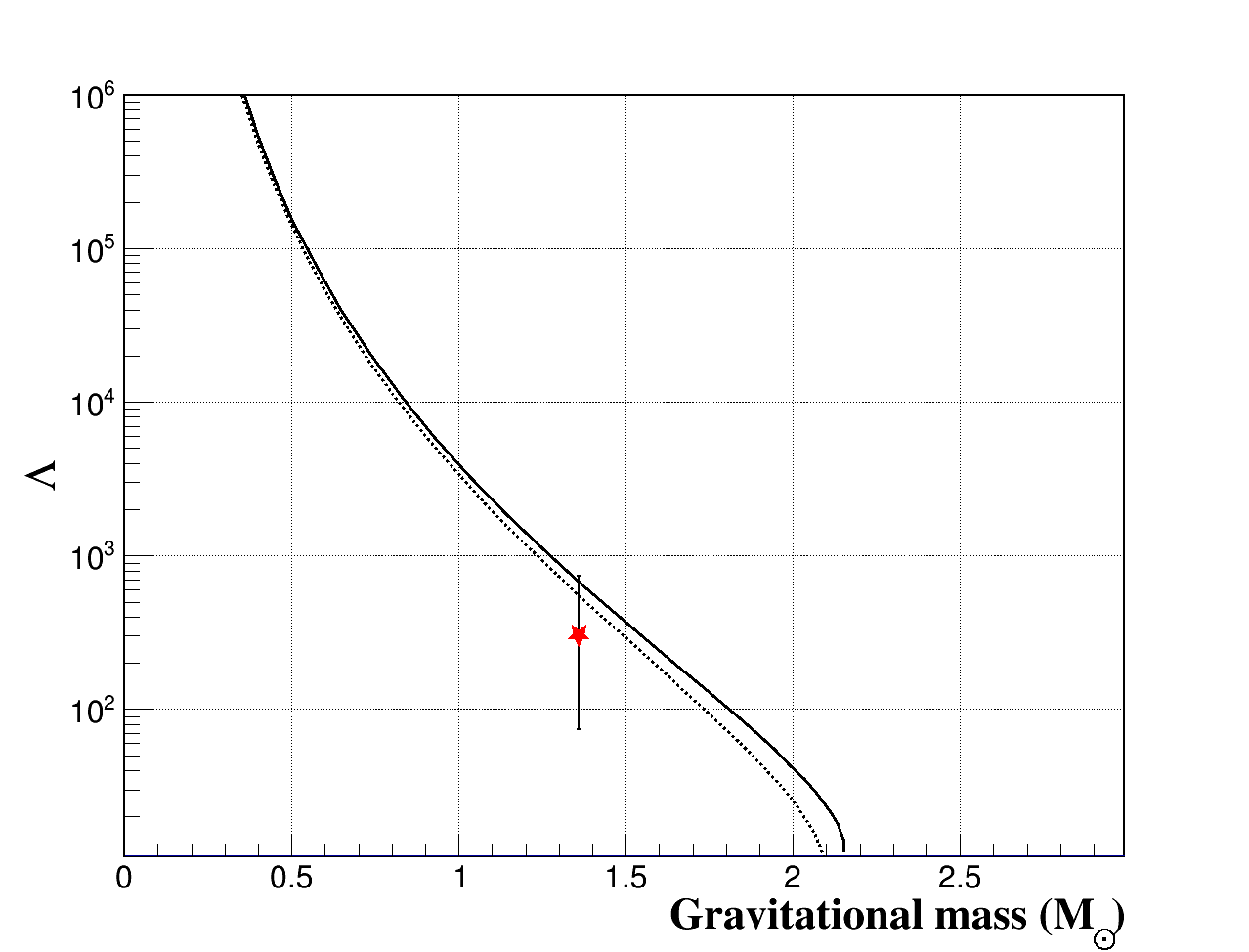}
\caption{(Left) Mass-radius diagram for m$_D$=400 MeV, $g_{vD}$/$m_{vD}$=0.015 MeV$^{-1}$ (solid line) compared to initial EOS from Table \ref{tab:tabeos} 
(dashed line). (Right) Tidal deformability for m$_D$=400 MeV, $g_{vD}$/$m_{vD}$=0.015 MeV$^{-1}$ (solid line) compared to initial EOS from Table \ref{tab:tabeos} (dashed line) and to constraint from binary NS merger GW170817 (star symbol). }
\label{fig4}
\end{figure*}

EOS which lead to reasonable results for mass-radius dependence were further used for calculation of tidal deformability. 
In Figure \ref{fig4} are shown mass-radius diagram and tidal deformability for m$_D$=400 MeV, $g_{vD}$/$m_{vD}$=0.015 MeV$^{-1}$ (solid line) compared to initial EOS from Table \ref{tab:tabeos}(dashed line) and to constraint from binary NS merger GW170817 (symbol). This was the most successful case, thus further strengthening the constraints on m$_D$ and $g_{vD}$/$m_{vD}$. 

The above analysis sets rather strong constraints to the mass and vector coupling of DM particle in present scenario based on most recent experimental results. There exist experimental constraints on mass and coupling of DM particles in given mass range. Recent measurement done by DAMIC-M collaboration \cite{DAMIC-M} presents rather strong constraints on interaction with electron via mixing of photon with dark photon. Also there exist astronomical constraints on the $g_{vD}$/$m_{vD}$, with typical value 4 -10 fm$^{2}$ \cite{Das-65,Das-66,Das-67,Das-68,Das-69} what corresponds to the values obtained here when converted into natural units. When applying the strictest constraint of 4 fm$^{2}$, corresponding to 0.01 MeV$^{-1}$, DM particle masses get restricted to 370-400 MeV. The existence of DM particles is still an open question and the specific scenario presented here depends on present status of experimental investigations of neutron decay puzzle which can evolve with the time. Repeating the recent measurement \cite{2412.19519} paying attention also to coincident positively charged particles appears to be a first priority.

\section{Conclusions}
In this work we investigate properties of NSs considering the implications of recent experimental observation suggesting that neutron decay is always accompanied by emission of electron while in 1\% of cases proton is not emitted. We develop a scenario kinematically compatible with experimental observation, where neutron decay results in production of two DM particles of about half mass of neutron besides energetic positron produced in three-body decay and carrying away the positive charge. Subsequently we test properties of NSs with admixture of such particles, including mass-radius diagram and tidal deformability. DM particle is assumed to be fermion with repulsive interaction mediated by a dark vector boson. Constraints on mass and coupling to scalar field are obtained. The combinations of DM particle mass m$_D$=370-400 MeV and vector coupling $g_{vD}$/$m_{vD}$=0.01-0.015 MeV$^{-1}$ appear most successful, being able to satisfy all the applied constraints. Structure of the resulting compact object is modified to a dominantly dark star with DM at the surface, a shell of nucleonic matter around the nuclear saturation density and a core again formed by DM. 


\end{document}